\documentclass[
 amsmath,amssymb,
 aps,longbibliography,
twocolumn
]{revtex4-1}

\usepackage{graphicx}
\usepackage{dcolumn}
\usepackage{bm}

\begin{document}

\title{Disorder Influences the Quantum Critical Transport at a Superconductor to Insulator Transition}

\author{H. Q. Nguyen}
\affiliation{Department of Physics, Brown University, Providence, RI 02912}
\author{S. M. Hollen}
\affiliation{Department of Physics, Brown University, Providence, RI 02912}
\author{J. Shainline}
\affiliation{School of Engineering, Brown University, Providence, RI 02912}
\author{J. M. Xu}
\affiliation{School of Engineering, Brown University, Providence, RI 02912}
\author{J. M. Valles Jr.}
\affiliation{Department of Physics, Brown University, Providence, RI 02912}

\begin{abstract}
We isolated flux disorder effects on the transport at the critical point of the quantum magnetic field tuned Superconductor to Insulator transition (BSIT). The experiments employed films patterned into geometrically disordered hexagonal arrays.  Spatial variations in the flux per unit cell, which grow in a perpendicular magnetic field, constitute flux disorder.  The growth of flux disorder with magnetic field limited the number of BSITs exhibited by a single film due to flux matching effects. The critical metallic resistance at successive BSITs grew with flux disorder contrary to predictions of its universality.  These results open the door for controlled studies of disorder effects on the universality class of an ubiquitous quantum phase transition.
\end{abstract}

\maketitle
Transport phenomena near quantum critical points receive ongoing scrutiny.  Much attention comes from efforts to understand strange metal behavior in high $T_c$ \cite{varma2002PR,vucicevicPRL2015} and heavy fermion compounds \cite{stewartRMP2001,faulknerScience2010}.  Others who are developing string theory based techniques  to calculate many body properties of condensed matter in the strong coupling limit have focussed on quantum critical transport as well\cite{sachdevARCMP2012,chen2014universal,witczakNP14}. An interesting case is the superfluid to insulator transition in charged two dimensional systems such as superconducting films.  This Superconductor to Insulator transition (SIT) appears in numerous thin film systems \cite{Gantmakherreview,dobrosavljevic12} as a change from superconducting to insulating transport when a physical parameter such as thickness \cite{dynes1978two,havilandPRL1989} magnetic field \cite{HebardPRL90} or a gating field \cite{parendoPRL2005,cavigliaNature2008,bollingerN2011,allainNM2012} is tuned.  At the critical value of the tuning parameter, film resistances asymptote to a constant value in the zero frequency, low temperature limit.  The limiting critical resistance assumes a value near the quantum of resistance for electron pairs $R_c\sim\frac{h}{4e^2}=R_Q$ \cite{Gantmakherreview,dobrosavljevic12}. Despite the simplicity of this behavior, questions remain regarding its universality and the influence of disorder.  

Indeed, the critical resistance near a superconductor to insulator transition is at the heart of studies, both theoretical and experimental. Fisher \cite{FisherPRL90a} and collaborators \cite{chaPRB1991} provided the original arguments supporting the universality of $R_c$.  They focussed on bosonic systems presuming that fermion degrees of freedom in the form of unpaired electrons play no role. They argued that $R_c$ is likely to be constant in ordered systems within a universality class \cite{chaPRB1991} especially in magnetic field.   Disorder appeared to lead to an increase in $R_c$.  Numerical simulations \cite{herbut1997dual,kimPRB08} also implied that quenched disorder modifies the SIT and $R_c$.  On the other hand, recent Quantum Monte Carlo simulations of ordered \cite{witczakNP14} and disordered systems \cite{swansonPRX14} produced nearly equivalent values of $R_c$ suggesting only a weak disorder dependence.  Similarly, experiments produce conflicting results.  For indium oxide films, $R_c$ hovers around $R_c\simeq R_Q$  \cite{HebardPRL90,SambandamurthyEPL06,SteinerPRB2008,KopnovPRL12} for a large variation in microscopic disorder.  Studies of homogeneous amorphous bismuth films \cite{MarkovicPRB98}, however, show that $R_c$ grows with the normal state resistance, which is a similar measure of disorder.  In both cases $R_c$ assumes values that are about a factor of two higher than observed in the most ordered systems, micro-fabricated Josephson Junction Arrays (JJA) \cite{ZantPRL92,ChenPRB95}.  Such an elaborate picture reflects the variety of approaches and systems employed to study $R_c$. It suggests that deriving a new method for controllably varying disorder and selecting a thin film system with a bosonic SIT is crucial to making further experimental progress.  

Opportunities to carefully test quantum critical transport models of bosonic systems have arisen for thin film superconductors.  There are clear indications that boson degrees of freedom dominate the SITs of a number of them \cite{Gantmakherreview,dobrosavljevic12}.   Increasing resistivity or applying a magnetic field transforms superconducting transport into thermally activated insulating charge transport consistent with boson localization in for example InO$_x$ \cite{SambandamurthyPRL04,steinerPhysicaC2005}, TiN \cite{BaturinaPRL07}, and nanostructured amorphous Bi films \cite{nguyenPRL09,LinPhysica14}.  Magnetoresistance measurements showing Little-Parks oscillations \cite{StewScience07,KopnovPRL12} indicate that the bosons, Cooper pairs of electrons, remain intact across the transition. Tunneling experiments imply that the gap in the quasiparticle density of states persists into the insulating phase \cite{sacepePRL2008,shermanPRL2012} indicating that the fermionic degrees of freedom are not active at these SITs. Near the critical point, resistance data from the prototypical indium oxide film system show the expected scaling behavior \cite{HebardPRL90,OvadiaNP2013,SteinerPRB2008}. Thus, recent work has made it possible to focus on bosonic phenomena using well chosen superconducting films. 

There are challenges to isolating the effects of disorder as it can take many forms that simultaneously influence film behavior at the SIT \cite{GhosalPRL98,SambandamurthyPRL04,BaturinaPRL07,DubiNature07,BouadimNP11}. Models consider disorder as random variations in the electron potential \cite{FisherPRL90a,GhosalPRL98,DubiNature07}, or intersite coupling in lattice models \cite{swansonPRX14}, or in physical parameters of grains in granular models \cite{BeloborodovPRB06}. Each of these can lead to qualitative accounts of SIT phenomena, such as the transition \cite{Gantmakherreview}, the emergence of granular structure in the Cooper pair distribution \cite{SacepePRL08}, and the appearance of the giant peak in the magnetoresistance of the insulating phase \cite{SambandamurthyPRL04}.  Distinguishing the influences of each type of disorder, however, has been problematic. The common disorder parameter, the normal state sheet resistance $R_N$, can reflect any of the above forms as it depends on carrier density, impurity potential, and film morphology \cite{Gantmakherreview}.  An additional confounding effect is that Coulomb interactions also grow with $R_N$.  These repulsive interactions inhibit Cooper pair tunneling between superconducting islands like those found in granular films \cite{EkinciPRL1999} and microfabricated JJAs \cite{FazioPR01}.  They strengthen as the interisland resistances increase through $R_Q$ to drive Cooper pair localization and thus, an SIT. 
 
\begin{figure*}[ht]
\begin{center}
\includegraphics[width=6in,keepaspectratio]{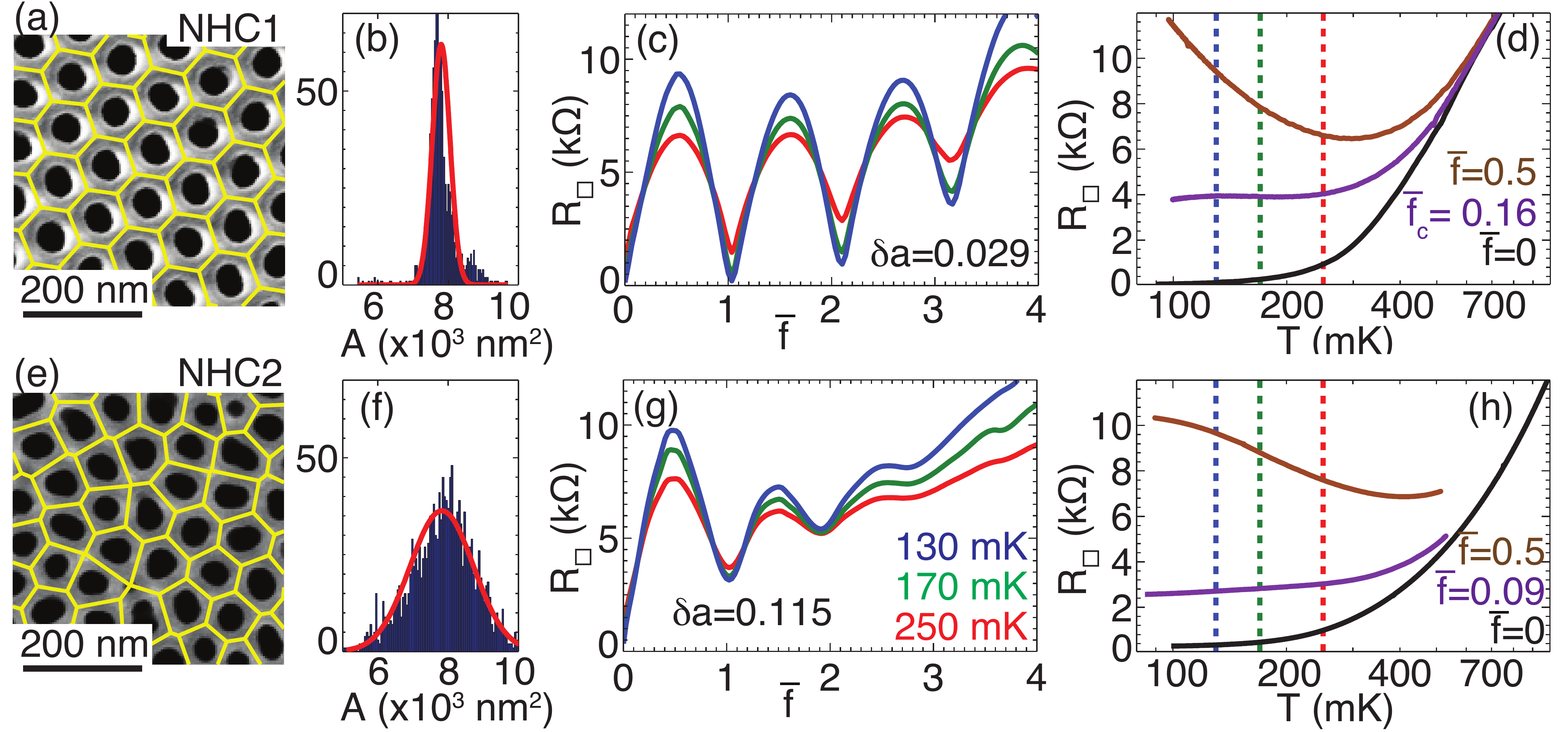}
\caption{Comparison of NHC films on two disordered arrays. (a-d) show data from the more ordered array NHC1 and (e-h) show data for the less ordered array NHC2.  (a, e) Electron micrographs of the substrates.  The unit cells, which were determined using a triangulation algorithm, are highlighted by yellow polygons. (b,f) Unit cell area distributions with Gaussian fits (red curves). The average unit cell area is $\bar A=8\times 10^3$ nm$^2$ for the two substrates. (c and g) Magnetoresistance oscillations ($\bar{f}=B\bar{A}/\Phi_0$) at 130, 170, and 250 mK. (d and h) Film resistances as a function of temperature at zero field ($\bar{f}=0$), near $f_c$, and at $\bar{f}=1/2$. The films on NHC1 and NHC2 have normal state resistances and thicknesses $R_N=17.9$ k$\Omega/\square$ and $d=1.22$ nm and $R_N=19.2$ k$\Omega/\square$ and $d=1.08$ nm, respectively.}
\label{disorder}
\end{center}
\end{figure*}

We developed a method to study quantum critical transport of a bosonic system near the SIT that isolates the effects of one form of disorder from other forms and interaction effects. We achieved this control by creating systems with tunable flux disorder. The method employs thin films patterned with a disordered triangular array of holes. The multiply connected geometry of these so-called nano-honeycomb (NHC) films enables a single film to exhibit a series of magnetic field driven bosonic SITs \cite{StewartPRB08} (see Fig. 1). The geometric disorder leads to variations in the number of flux quanta per unit cell. This flux disorder \cite{GranatoPRB86} grows with magnetic field so that successive SITs occur with increasing disorder. It limits the number of magnetic field tuned SITs that appear due to flux matching effects.  More notably, rather than being universal, the $R_c$ of these SITs increase with flux disorder from about 4 k$\Omega$ per square to plateau at about 6 k$\Omega/\square$.  We discuss how this observation implies that geometric array disorder presumed to exist in microscopically disordered thin films \cite{DubiNature07,SacepePRL08} enhances $R_c$ compared to ordered Josephson junction arrays. 

We used anodized aluminum oxide substrates with a nearly triangular array of holes \cite{YinAPL01} as a template for NHC films (Figs. 1a,e).  All substrates had the same average hole spacing of 100 nm.  The more strongly disordered were produced by wrapping the aluminum with teflon tape to perturb the normally laminar flows that set up during anodization \cite{YinAPL01}.  The geometric disorder is apparent in histograms of the unit cell areas for two typical NHC substrate (Figs. 1b,f).  The substrates were mounted to a dilution refrigerator and held at 8 K during film deposition.  We studied arrays with fixed geometric disorder and varying normal state sheet resistance by depositing a series of amorphous Bi films on a single substrate \cite{StewScience07,HollenPRB11}. Sheet resistances were measured at low frequencies using four probes. Transverse magnetic fields $B$ were applied using a superconducting solenoid and are specified by the average number of flux quanta per unit cell in the array  $\bar f=B \bar A/\Phi_0$, where $\bar A$ is the average unit cell area and $\Phi_0$ is a flux quantum. 

Flux disorder results from variations in the geometry of the network of the templated NHC films.  We characterize the network disorder by the fractional variation in the unit cell areas, $\delta a \equiv \Delta A / \bar A$ where $\Delta A$ is the standard deviation calculated from gaussian fits to the unit cell area histograms (Figs. 1 b,f).  In a magnetic field, there are variations in the local frustration, $\delta f = \bar f\delta a$ that constitute the flux disorder \cite{GranatoPRB86}.  This linear growth of $\delta f$ with magnetic field is presumed to dominate any field induced changes in other forms of disorder.  In particular, previous work by our group \cite{HollenPRB11} demonstrated that NHC films are comprised of an array of islands connected by weak links.  It is supposed that randomness in weak link coupling and island size varies little with magnetic fields well below the estimated upper critical magnetic field\cite{Nguyen2010}. 

The effects of flux disorder can be seen in a comparison of the low temperature magneto transport of two films with different geometric disorder but similar $R_N$  (Figs. 1 c,g).   In both cases, the magnetoresistance oscillates with a period of 1 between low values at integer $\bar f$ and high values at half-integer $\bar f$ \cite{nguyenPRL09}).  The oscillations decay more rapidly with $\bar f$ for the more disordered sample. Investigations of multiple substrates\cite{Nguyen2010} indicate that the number of visible oscillations decreases from about 5 to 1 as $\delta a$ increases from 0.03 to 0.14 and does not depend on $R_N$\cite{Supp}.  Thus, the data imply that oscillations persist only up to fields such that the flux disorder $\delta f\approx 0.3$. The maximum number of oscillations observed in the most ordered arrays appears to be limited by the rise in the magneto-resistance that develops beyond 1 Tesla \cite{nguyenPRL09}.

Insight into the origin of the oscillations described above comes from prior experiments by Forrester et. al. \cite{BenzPRB88,ForresterPRB88} on micro-fabricated Josephson junction arrays with geometrical disorder.  Their magneto-resistances near the superconducting transition temperature $T_c$ oscillated with a period of 1 and decayed more rapidly in more disordered arrays. The oscillations completely disappeared above a critical field that was inversely proportional to the amount of geometrical disorder.  The visibility of just 3 oscillations in Fig. 1g for $\delta a$ = 0.115 is in rough accord with their results.  In addition, they showed that the phenomenon results from $T_c$ variations caused by oscillations of the average of the Josephson coupling energy $E_J=<-J \cos(\theta_i-\theta_j-A_{ij})>$ between neighboring nodes in the array \cite{BenzPRB88}.  $J$ is proportional to the amplitude of the superconducting order parameter on the nodes,  $\theta_i-\theta_j$ is the difference in the phases of the order parameter, and $A_{ij}=\frac{h}{2e}\int_i^j \bf A\cdot dl$ is the line integral of the magnetic vector potential between neighboring nodes.   $E_J$ oscillates as $A_{ij}$ increases with magnetic field and the oscillation amplitude decays as the variations in $A_{ij}$ grow. The resemblance of the oscillations presented in Fig. 1 with the prior results indicates that a similar modulation of $E_J$ occurs in the NHC films.  Its effect, however, appears more dramatic as near the SIT, $E_J$ controls the strength of quantum phase fluctuations that potentially drive NHC films into an insulating state.  

Quantum superconductor to insulator transitions \cite{GranatoPRB13} are evident in the magneto-resistance oscillations in Figs. 1c,g. They are identified by the crossing points in the $R_{\square}(B)$ traces taken at different temperatures. At each critical point $(\bar f_c,R_c)$, the slope $\frac{dR_{\square}}{dT}$ changes sign \cite{StewartPRB08} as depicted by the $R_{\square}(T)$ in Figs. 1d,h. Negative and positive slopes correspond to films on the insulating and superconducting sides of the SIT, respectively (cf. Ref. \cite{Gantmakherreview}). Changes in the slopes with field appear to be governed by smooth changes in an activation energy\cite{StewartPRB08}, which is consistent with the onset of boson localization.  Also, the insulating film $R_{\square}(T)$s show reentrant dips indicative of Cooper pairing fluctuations \cite{StewScience07} prior to rising at low temperatures (Figs. 1d,h). Seven crossing points are apparent in the more ordered sample while only three appear in the less ordered sample.   

\begin{figure}[ht]
\begin{center}
\includegraphics[width=3in,keepaspectratio]{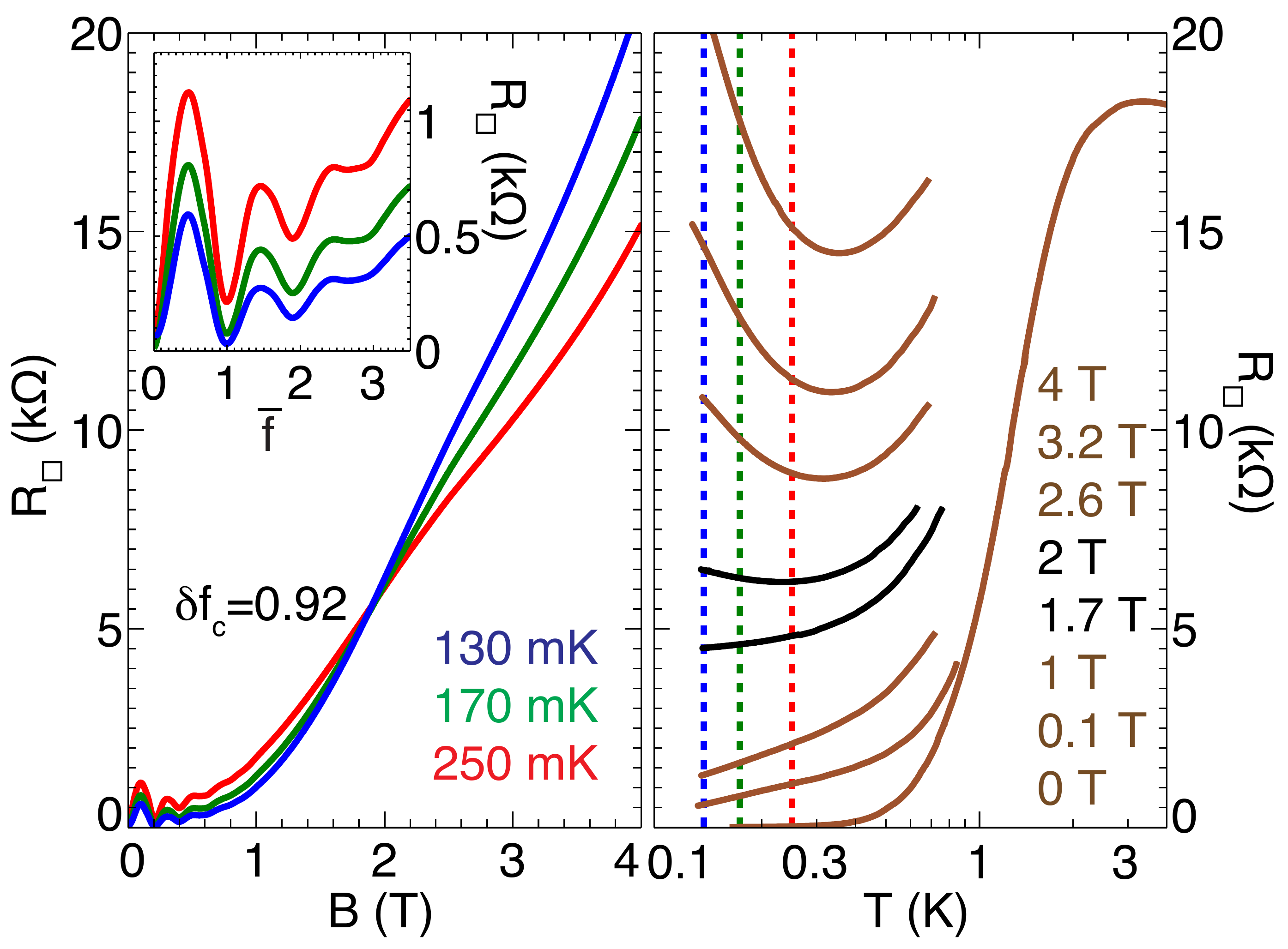}
\caption{A high field SIT. Film on NHC2 with $R_N=17$ k$\Omega/\square$.  (a) Isothermal magnetoresistance curves at 130, 170, and 250 mK show a single crossing at $B_c=1.9$ T. Inset:  Expanded low field region.  (b) $R_{\square}(T)$ at discrete magnetic fields spanning $B_c$. Vertical dashed lines indicate the isothermal slices in (a)}
\label{highBSIT}
\end{center}
\end{figure}

It is also possible to observe a bosonic superconductor to insulator transition at fields beyond the oscillation regime as shown in Fig. 2.  The overlay of $R_{\square}(B)$ traces at different temperatures shows a crossing near 1.9 T. This critical field corresponds to $\bar f_c\approx 8$ or $\delta f\approx 0.9$. Qualitatively, $R_{\square}(T)$s at fixed magnetic fields develop tails at low temperatures that evolve into a flat dependence at the critical point with $R_c\approx$ 6 k$\Omega/\square$ and finally an upturn with increasing field. These positive slopes $dR_{\square}/dT$ fit to an Arrhenius form, which implies an activation energy of these insulating films \cite{StewartPRB08}. Also, these $R_{\square}(T)$s show Cooper pairing reentrant dips. This evolution resembles the SITs in the oscillation regime of thinner films (see Figs. 1d, h) intimating that this high field SIT is also bosonic.  The magnetic field induced reduction of $E_J$ that drives this transition, however, must occur via a decrease in $J$ caused by pairbreaking effects rather than an increase in frustration $\sim A_{ij}$.    

The quantum critical resistances change systematically with flux disorder as illustrated in Fig. 3.  The $R_c$ obtained from films shown here and others \cite{Nguyen2010} were determined from crossing points in continuous field sweeps as in Fig. 1 or by interpolating measurements of $R_{\square}(T)$ and $\frac{dR_{\square}}{dT}$ at discrete fields to $\frac{dR_{\square}}{dT}=0$ \cite{StewartPRB08}.  These methods yielded similar results when they could be compared.  It is apparent in the inset that $R_c$  increases with flux in the low flux limit for fixed geometric disorder.  Linear fits to data from two individual films emphasize this monotonic rise.  This behavior contrasts with ideal JJA's for which $R_c$ is independent of flux.  Plotting the $R_c$ of a number of films versus flux disorder $\delta f$ reveals that $R_c$ rises with a slope of $\sim 3-4.5$ k$\Omega/\square$ per unit of flux disorder to saturate near $\frac{h}{4e^2}$ for $\delta f\geq 0.3$.  $R_c(\delta f\rightarrow 0)$ varies from 2.5 to 4 $k\Omega$.  Experiments on 1 set of films suggest that $R_c(\delta f\rightarrow 0)$ depends on weak link coupling, which is unexpected for ideal JJA's \cite{Supp}.

The flux disorder dependence of $R_c$ in Fig. 3 provides an explanation for the difference in $R_c$ measured in films and fabricated JJAs.  Films showing bosonic SIT characteristics exhibit $R_c\approx R_Q$ with no clear dependence on $R_N$.  Most notably, data on many different InO$_x$ films indicate $R_c\approx 5.8$ k$\Omega$ \cite{HebardPRL90,SambandamurthyEPL06,KopnovPRL12}. Because these films lack clear geometrical structure, their BSITs presumably occur in the high flux disorder limit.  Experiments on JJAs, on the other hand, yielded $2.5<R_c<  4.5$ k$\Omega$ \cite{ZantPRL92} and $1.2<R_c< 2.45$ k$\Omega$ \cite{ChenPRB95}. Clearly BSITs occur in the zero flux disorder limit in JJAs.  Altogether these experiments point to $R_c$ increasing with flux disorder (cf. Fig. 3).   

\begin{figure}[ht]
\begin{center}
\includegraphics[width=2.8in,keepaspectratio]{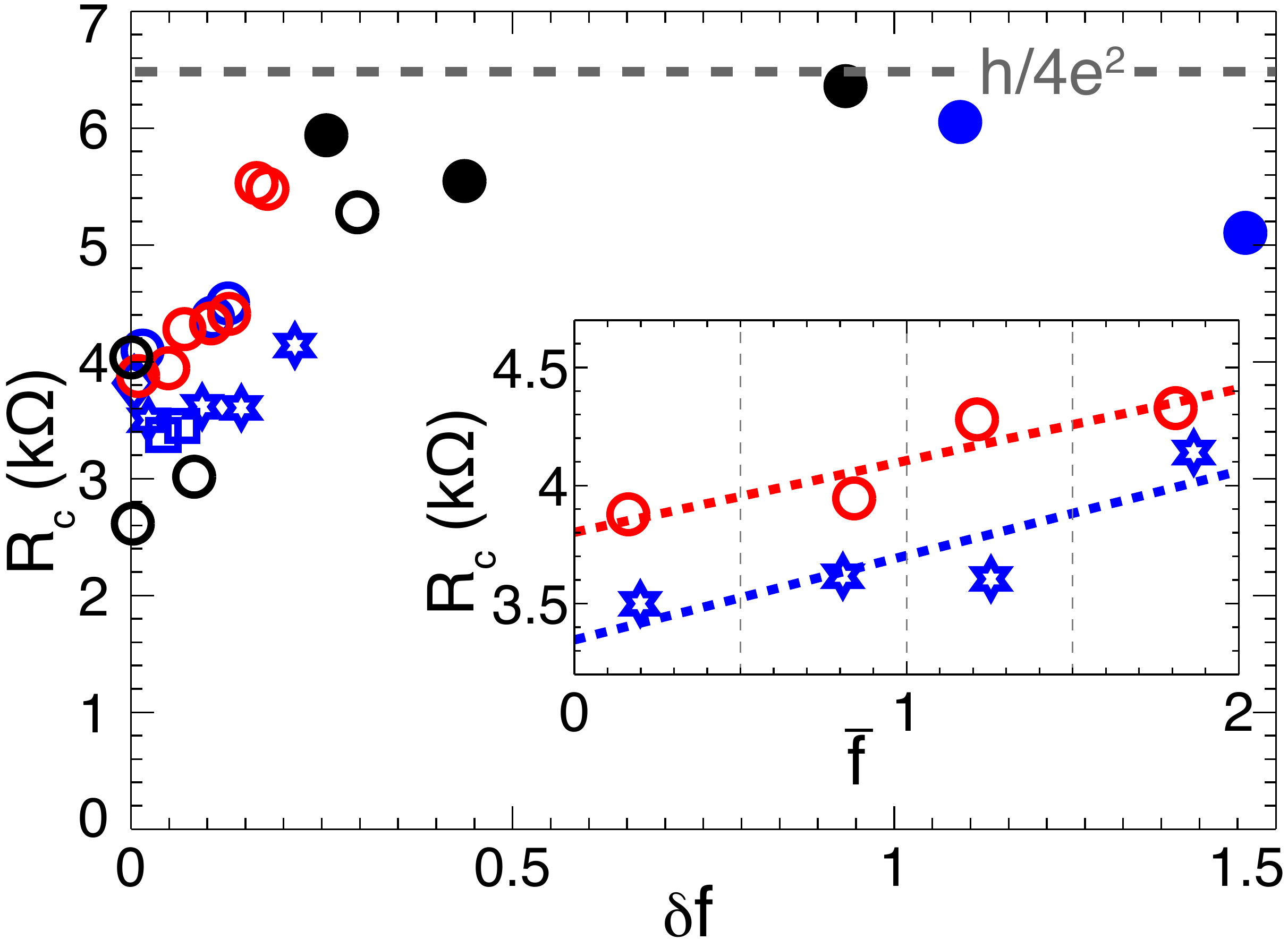}
\caption{Critical resistance as a function of flux disorder. Closed circles are high field SITs. Open symbols are SITs within the MR oscillation regime, with red circles from NHC1 ($R_N = 19.7$ k$\Omega/\square$) and blue symbols from NHC2: Diamonds ($R_N = 20.3$ k$\Omega/\square$), stars ($R_N = 19.2$ k$\Omega/\square$), open circles ($R_N = 17.4$ k$\Omega/\square$), squares ($R_N = 16.7$ k$\Omega/\square$), solid circles ($R_N = 16$ k$\Omega/\square$). Black symbols are from other substrates. Inset: $R_c$ as a function of $\bar f$ for two of the films. Dashed lines are linear fits to the open circles of the same color.}
\label{HcRc}
\end{center}
\end{figure}

It is important to view the current results relative to previous experiments showing a relation between $R_c$ and disorder as parameterized by the normal state sheet resistance $R_N$.  Investigations of homogeneous MoGe films \cite{YazdaniPRL95} and uniform amorphous Bi films \cite{MarkovicPRB98} showed a strong positive correlation between $R_c$ and $R_N$.  This correlation could imply that $R_c$ increases with disorder \cite{MarkovicPRB98}.  Alternatively, it could be produced by fermion quasiparticles known to be present at these SITs \cite{HsuPRL1995}.  These fermions can provide a parallel dissipative channel that alters the transport in the vicinity of a nominally bosonic quantum critical point \cite{masonPRL1999}.  Thus, the correlation between $R_c$ and $R_N$ could reflect changes in quasiparticle density rather than disorder \cite{YazdaniPRL95}.  We hasten to add that patterning these films with hole arrays can potentially yield insight into the interplay of bosonic and fermionic degrees of freedom at an SIT.  It can reveal how dissipation effects on a bosonic SIT due to quasiparticles \cite{masonPRL1999} evolve with increasing disorder.   

Although there have been many calculations of $R_c$ at bosonic superconductor to insulator transitions including some that explicitly consider disorder effects \cite{FisherPRL90a,chaPRB94,herbut1997dual,nishiyamaPhysC01,swansonPRX14,kimPRB08,wallinPRB1994} none appear to predict the behavior observed here.  Uniquely, Kim and Stroud \cite{kimPRB08} treated flux disorder effects in Quantum Monte Carlo simulations of disordered square arrays of Josephson junctions.  Their results predict that $R_c$ decreases strongly with flux disorder from $3.5 R_Q$ to $0.2R_Q$ for integer $\bar f$, which is in sharp contrast to our results at non-integer $\bar f$ in Fig. 3.  Likewise, a comparison \cite{witczakNP14} of a pair of Quantum Monte Carlo results \cite{swansonPRX14,witczakNP14} employing the latest methods for extracting $R_c$ from simulations suggests that $R_c$ only weakly depends on disorder.  In addition, the vast majority of predictions of $R_c$ in the low disorder limit are higher than experimental values by a factor of more than two \cite{BatrouniPRL93,SorensenPRL92,MakivicPRL93,RungePRL92,swansonPRX14,witczakNP14,herbut1997dual,chen2014universal}. This discrepancy could suggest that the universality classes of the quantum rotor and Villain models employed in some of the most advanced approaches \cite{swansonPRX14,witczakNP14} do not match the experimental systems.  Or, it could reflect a need to further refine methods for calculating the quantum critical conductivity in the zero frequency limit \cite{swansonPRX14,witczakNP14,damlePRB1997}.  There is the further possibility that experiments have not probed the true quantum critical regime.    

In conclusion, we have described experiments revealing that disorder influences quantum critical transport of a bosonic system.  We find that the critical resistance at the superconductor to insulator transition depends on flux disorder.  This result invites more theoretical attention to its universality while illuminating a difference between SITs in thin films and Josephson junction arrays.  The results also highlight a lack of quantitative agreement between theoretical predictions and measurements of the critical resistance. The studies establish NHC films as uniquely suited for studying  the effects of well defined disorder on quantum critical transport and the universality class of a prominent quantum phase transition. Further, they invite comparisons with other condensed matter systems in which the interplay of quantum criticality and disorder can arise such as doped strongly correlated electron systems \cite{roschPRL1999} or cold atoms in random optical lattices \cite{deisslerNP2010}. 

\begin{acknowledgments} 
We have benefitted from discussions with E. Granato, A. Frydman, J. Joy, Xue Zhang and N. Trivedi. We are grateful for the support of NSF Grants No. DMR-1307290 and No. DMR-0907357 and AFOSR and AOARD.
\end{acknowledgments}

\bibliography{rcdisorder}
\end{document}